\documentclass{birkjour}
 \newtheorem{thm}{Theorem}[section]

 \theoremstyle{definition}
 
 \theoremstyle{remark}
 \newtheorem{rem}[thm]{Remark}
 
 \numberwithin{equation}{section}
\usepackage{url}

\begin{document}
\title[Analysis of the COVID-19 by SIR model and forecasting]
 {Analysis of the   COVID-19 pandemic by SIR model and machine learning technics for forecasting}

\author[Babacar et al.]{Babacar Mbaye Ndiaye, Lena Tendeng, Diaraf Seck}
\address{Laboratory of Mathematics of Decision and Numerical Analysis\br
University of Cheikh Anta Diop\br
BP 45087, 10700. Dakar, Senegal}

\email{babacarm.ndiaye@ucad.edu.sn, lenatendeng@yahoo.fr,\\ diaraf.seck@ucad.edu.sn}

\thanks{This work was completed with the support of the NLAGA project}


\date{April 1st, 2020}
\begin{abstract}
This work is a trial in which we propose SIR model and machine learning tools to analyze the coronavirus pandemic in the real world. Based on the public data from \cite{datahub}, we estimate main key pandemic parameters and make predictions on the inflection point and possible ending time for the real world and specifically for Senegal.   
The coronavirus disease 2019,  by World Health Organization, rapidly spread out in the whole China and then in the whole world.
Under optimistic estimation, the pandemic in some countries will end soon,  while for most part of countries in the world (US, Italy, etc.), the hit of anti-pandemic will be no later than the end of April. 
\end{abstract}

\maketitle
\section{Introduction}
\begin{figure}[h!]
	\centering
	\includegraphics[width=1.1\linewidth]{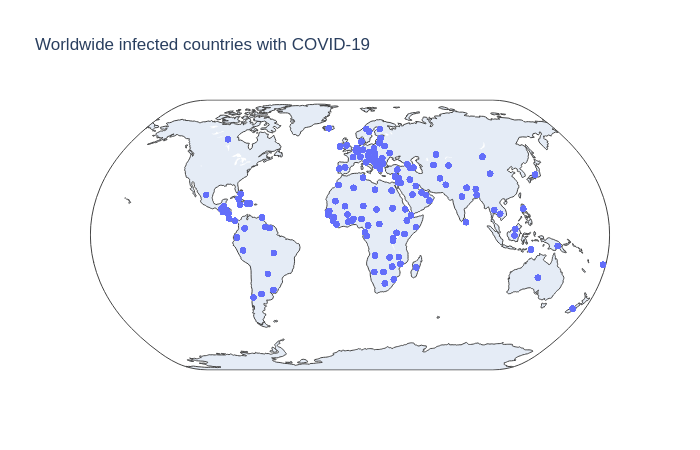}
	\par\vspace{-1.5cm}
	\caption{All infected countries with COVID-19 pandemic} 	\label{allinfected}
\end{figure}

\noindent The pandemic of COVID-19 was firstly reported in Wuhan in December 2019 and then quickly spread out international wide, see Figure \ref{allinfected}.\\
Since some decades ago mathematical community and specialists in computer sciences investigate  biological questions to propose models  describing  subjects such as diseases, physiological phenomena. One of these models are formulated  with the help of  Differential Equations (SIR models, reaction diffusions equations, etc.) while others use computer sciences (see \cite{steven,norden,Kamalich,hiroshi,earn}).\\
In this work, for the mathematical modeling part, we are going to  use the SIR model without demography because the lifespan of the disease which is an influenza disease will be very short compared to the lifespan of humans (even if the flu lasted 2 years which would be exceptional, 2 years do not represent much compared to the average longevity of a man). Similarly for this type of disease it is also assumed that there will be no significant renewal of the population while it is raging. The last influenzas that seriously affected humanity did not last long (H1N1 for example) even if this new virus hitherto seriously defies science.\\
\noindent Our aims are first, to collect carefully the pandemic data from the \cite{datahub},  
e.g. \url{https://www.tableau.com/covid-19-coronavirus-data-resources},  from January 21, 2020 to April 02, 2020. In the second step,  we propose two technics,  a  SIR model and machine learning tools, to analyze the coronavirus pandemic in the worldwide.
\par\vspace{-.25cm}
\noindent The article is organized as follows. In the section \ref{aux}, we present some auxiliary results of the SIR model (model, estimation of parameters) and the machine learning technics for forecasting, followed by the simulations and results in section \ref{SimNumerical}. Finally, in the section \ref{ccl}, we present conclusions and perspectives.

\section{Methodology}\label{aux}
\subsection{SIR model}
In this work, we use the classical Kermack-Mckendrick SIR model to describe the transmission of the COVID-19 virus pandemic\cite{Kermack,Kermack2}.
The SIR model is a compartmental model for modeling how a disease spreads through a population. It’s an acronym for Susceptible, Infected, Recovered.  The total population is assumed to be constant and divided into three classes.
\begin{figure}[h!]
	\centering
	\includegraphics[width=.8\linewidth]{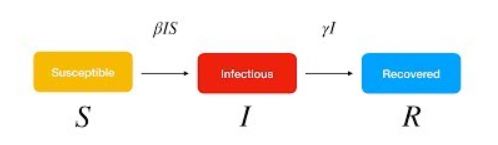}
		\par\vspace{-.5cm}
	\caption{SIR model diagram}  \label{deathsforcast}
\end{figure}

\noindent
In this model, naturel death and birth rate are not considered because the outstanding period of the disease is much shorter than the lifetime of the human.
\begin{equation}\label{}
\left\{ 
\begin{array}{ll}
\frac{dS}{dt}=-\beta IS\\[3mm]
\frac{dI}{dt}= \beta IS-\gamma I\\[3mm]
\frac{dR}{dt}=\gamma I
\end{array}\right.
\end{equation}
\begin{itemize}
\item $S$ is the number of individuals susceptible to be infected at time $t$.
\item $I$ is the number of both asymtotimatic and symptomatic infected individuals at time $t$.
\item $R$ is the number of recovered persons at time $t$.
\item The parameters $\beta$ and $\gamma$ are the transmission rate through exposure of the disease and the rate of recovering.
\end{itemize}
Fore more details about SIR, see \cite{howard}.

 \subsection{Estimation of parameters}
Parameters identification methods fall into two broad categories: parametric and non-parametric. The aim of non-parametric methods is to determine models by direct technics, without establishing a class of models a priori. They are called nonparametric because they do not involve a vector of parameters to be sought to represent the model.\\
Parametric methods are based on model structures chosen a priori and parameterized. In this case, the goal is to find a vector of parameters $\theta$, which will make it possible to obtain a model representing the behavior as close as possible to the real system. The search for the best model becomes a problem of estimating $\theta.$\\
Parameter identification belongs to the  topic of inverse  problems\cite{NLS,SaSe,T}. It is widely  studied and there is an abundant literature. But the subject is still interesting and there are still open problems. There are numerous  approaches for  the estimation of parameters.
In this paper, we shall  give a brief overview and quote  the least square method. For details and the discovery of other methods see for instance \cite{T}.\\
If we look at the problem in general first, we have data $y_i, i= 1, ..., n$ where each $y_i$ was measured at a specific $t_i$ point; on the other hand, we have a model, i.e. a function $f$ depending on time on the one hand and on the other hand a set of parameters that will be noted $P= (p_1, p_2,...,p_k)$. It is assumed that the amount (scalar or vector) that is tracked is described by $q(t)= f (t, p_1, p_2,..., p_k)$ that is   noted by $f (t, P)$. We do not  know the value of $p_j, j= 1,... ,k$ but with measurement errors, we must have the relationship $$y_i= f (t_i, P), \quad i= 1,..., n$$
   and then that $ y_i$ is the measure of $q(t_i)$. \\
   The idea is therefore to look for all the values $p_1, p_2,..., p_k$, which, in a sense, makes the differences $y_i-f (t_i, P)$  as close as possible to $0$ for all $i=1,..., n $ simultaneously.\\
 One natural method is the least square one, by setting $$\displaystyle \mathcal E(P)= \sum_{i=1}^n [ y_i-f (t_i, P)]^2$$ to be minimized on a subset $\mathcal{S}$ of $\mathbb{R}^k.$ This set may be open of closed. Even it is quite possible to study cases where Euclidean metric is replace by a Riemannian one usually noted $g.$
\begin{rem}
To estimate parameters,  it is possible to use the observer theory by first ensuring that the model is well observable and identifiable.
\end{rem}
\subsection{Existence}
First let us remind that the  minimization problems in the previous section are  very  classical in the cases we shall consider. That is why, we do not emphasize on the  theoretical aspects.\\
For  the SIR model, it is also classical to have existence and uniqueness whenever convenient Cauchy data are considered.  The Cauchy Lipschitz theorem ensure existence and uniqueness of local solution. And  even the recovery theorem is equivalent to  Cauchy Lipschitz theorem.

\subsection{Forecasting using Prophet}
Machine learning technics for forecasting is a branch of computer science where algorithms learn from data. Algorithms can include artificial neural networks, deep learning, association rules, decision trees, reinforcement learning and bayesian networks \cite{arkes,armst,litsa}.\\
Forecasting involves taking models fit on historical data and using them to predict future observations. We use this technic for the COVID-19 prediction in the real world.\\
We use Prophet \cite{prophet, sean}, a procedure for forecasting time series data based on an additive model where non-linear trends are fit with yearly, weekly, and daily seasonality, plus holiday effects. It works best with time series that have strong seasonal effects and several seasons of historical data. Prophet is robust to missing data and shifts in the trend, and typically handles outliers well. \\
For the average method, the forecasts of all future values are equal to the average (or “mean”) of the historical data. If we let the historical data be denoted by $y_1,...,y_T$, then we can write the forecasts as
$$
\hat{y}_{T+h|T}=\bar{y}=(y_1+y_2+...+y_T)/T
$$
The notation $\hat{y}_{T+h|T}$ is a short-hand for the estimate of $y_{T+h}$  based on the data $y_1,...,y_T$.\\
A prediction interval gives an interval within which we expect $y_t$  to lie with a specified probability. For example, assuming that the forecast errors are normally distributed, a 95\% prediction interval for the  $h$-step forecast is 
$$
\hat{y}_{T+h|T}\pm1.96\hat{\sigma_h}
$$
where  ${\sigma_h}$ is an estimate of the standard deviation of the $h$-step  forecast distribution.  \\
We perform one week ahead forecast with Prophet, with 97\% prediction intervals (multiplier=2.17). Here, no tweaking of seasonality-related parameters and additional regressors are performed.\\
For the data preparation, when we are forecasting at country level, for small values (Senegal for example), it is possible for forecasts to become negative. To counter this, we round negative values to zero.

\section{Numerical simulations and results}\label{SimNumerical}
In this section, we present the simulations of SIR model and the forecating technic for the prediction of the pandemic. \\
The numerical tests were performed by using the Python with panda \cite{python}. The numerical experiments were executed on a computer intel(R) Core-i7 CPU 2.60GHz, 24.0Gb of RAM, under UNIX system.

\subsection{Data analysis}
\begin{figure}[h!]
	\centering
	\includegraphics[width=.8\linewidth]{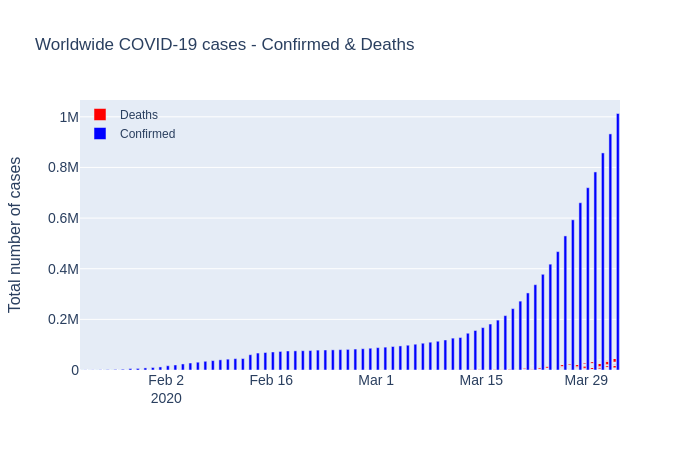}
		\par\vspace{-1.cm}
	\caption{Worldwide cases - confirmed and deaths} 	\label{cd_c}
\end{figure}

\begin{figure}[h!]
	\centering
	\includegraphics[width=.8\linewidth]{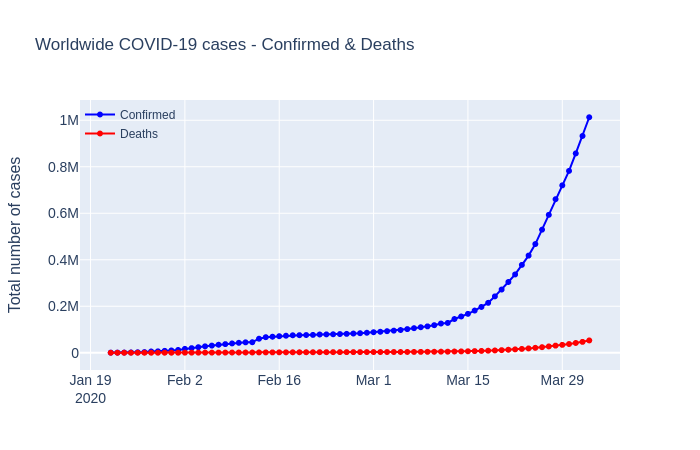}
		\par\vspace{-1.cm}
	\caption{Worldwide cases - confirmed and deaths} \label{cd_b}
\end{figure}

As cited in the introduction, the simulations are carried out from  data in \cite{datahub} (\url{https://www.tableau.com/covid-19-coronavirus-data-resources}), from January 21, 2020 to April 02, 2020. 
According to this daily reports, we first analyse and make some data preprocessing before simulations. The cumulative numbers of confirmed cases and deaths cases are illustrated in Figure \ref{cd_c} and Figure \ref{cd_b} (curve and histogram). 

\subsection{SIR model}
We use the SIR model simulations for the Senegal case study. For the initial values, we’ll use normalized population values for our $S_0$, $R_0$, etc. So if we assume we have 10k people in our population, and we begin with four  exposed person and the remaining susceptible, we have:
$N=10k$ (a sample), $I_0 = 4, R_0 = 0, S_0 = N-I_0-R_0$. \\
For the contact rate ($\beta$) and the mean recovery rate ($\gamma$), we have: $\beta=0.25$ and $\gamma=1/30$ (in 1/days).
\begin{figure}[h!]
	\centering
	\includegraphics[width=.85\linewidth]{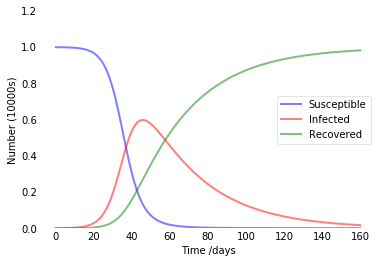}
	\caption{Prediction with SIR for the Senegal} 	\label{fig_sir}
\end{figure}
\noindent Susceptible people in blue curve remain constant for less than twenty days which could correspond to the incubation periode of the disease.\\
After the day 20, there is a rapid decrease in the blue curve which show the fast transmission of COVID-19.\\
Infected people in red line appear right after day 20. Where the red curve and blue curve cross, at about day 40, is the first day where more people are in the infected compartement  than the susceptible one.\\
The peak of infected people is reached at day 40 approximately with 6000 infected persons (60\% of infected persons). The model predicts a significant number of recoverers so few deaths due to the disesase.

\subsection{Forecast for the pandemic of COVID-19}
\begin{figure}[h!]
	\centering
	\includegraphics[width=.9\linewidth]{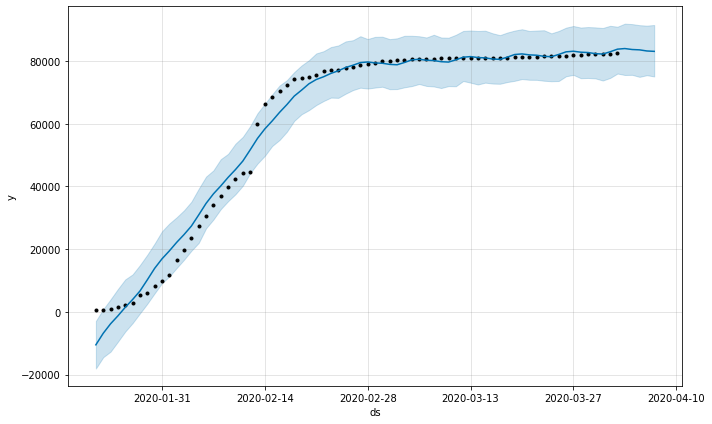}
	\caption{Forcasting - confirmed in China}  \label{ch}
\end{figure}

\begin{figure}[h!]
	\centering
	\includegraphics[width=.9\linewidth]{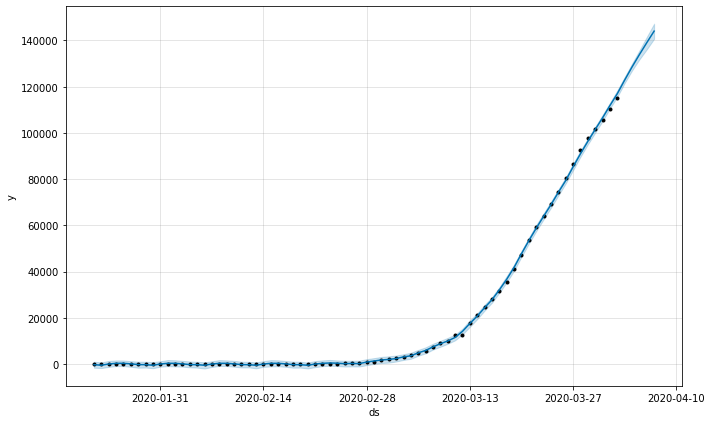}
	\caption{Forcasting - confirmed in Italy}  \label{it}
\end{figure}

\begin{figure}[h!]
	\centering
	\includegraphics[width=0.9\linewidth]{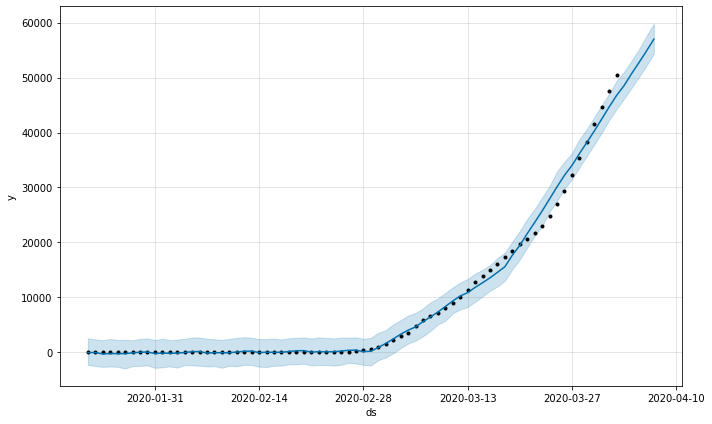}
	\caption{Forcasting - confirmed in Iran}  \label{ir}
\end{figure}

\begin{figure}[h!]
	\centering
	\includegraphics[width=0.9\linewidth]{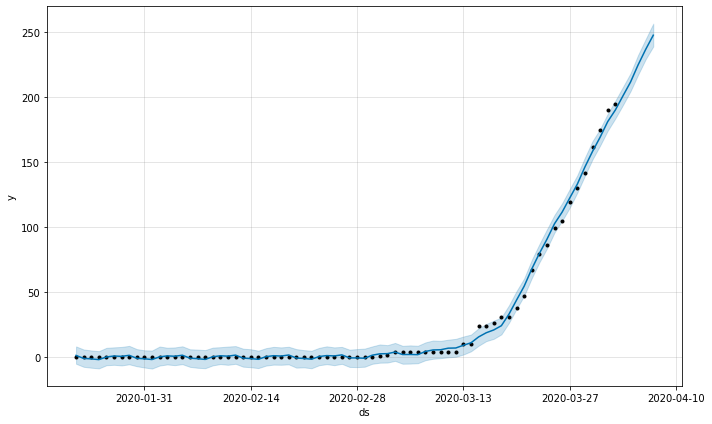}
	\caption{Forcasting - confirmed in Senegal}  \label{sn}
\end{figure}

Here, we perform predictions for the worldwide (180 countries, from January 21, 2020 to April 02, 2020) and four selected countries China, Italy, Iran and Senegal.
\begin{table}[h!] 
\begin{center}
\begin{tabular}{|c|c|c|c| } 
 \hline
          ds   &         $ \hat{y}$ &    $\hat{y}_{lower}$ &    $\hat{y}_{upper}$ \\
           \hline
2020-04-03  &	9.740387e+05 	&9.259386e+05 &	1.017860e+06\\
2020-04-04  &	1.024429e+06 &	9.780116e+05 &	1.072039e+06\\
2020-04-05  &	1.074312e+06 &	1.029720e+06 	&1.123804e+06\\
2020-04-06  &	1.125356e+06 	&1.074945e+06 	&1.172525e+06\\
2020-04-07  &	1.177793e+06& 	1.129130e+06 	&1.229721e+06\\
\hline
\end{tabular}
\end{center}
\caption{Predicted cumulative confirmed cases in the near future 
$\sim$April 07, 2020.}\label{futurconfcases}
\end{table} 
\begin{figure}[h!]
	\centering
	\includegraphics[width=.9\linewidth]{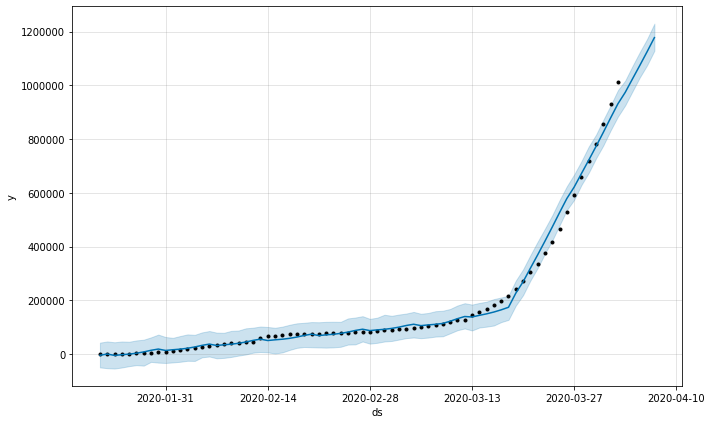}
	\caption{Forcasting - confirmed cases in the worldwide} 	\label{ww_conf}
\end{figure}
\begin{figure}[h!]
	\centering
	\includegraphics[width=.9\linewidth]{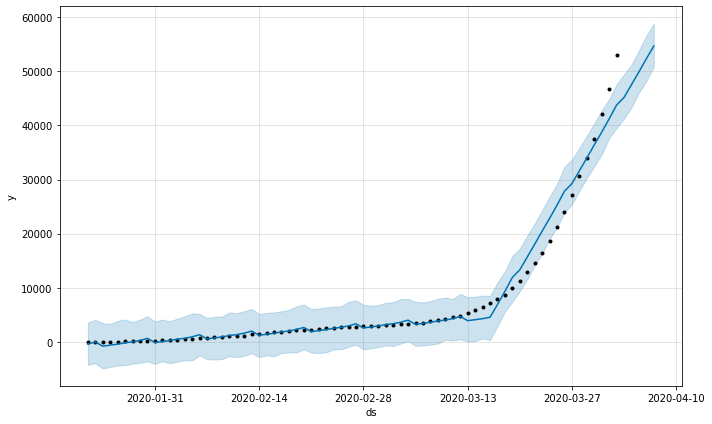}
	\caption{Forcasting - deaths cases in the worldwide}  \label{ww_dt}
\end{figure}
\noindent Most importantly, with the model and parameters in hand, we can carry out simulations for a longer time and forecast the potential tendency of the COVID-19 pandemic. For all figures, the predicted cumulative number of confirmed cases and the number of deaths cases are plotted for a shorter period of next 5 days.\\
Under optimistic estimation, the pandemic in some countries (like China) will end soon within few weeks,  while for most part of countries in the world (Italy, Iran and Senegal), the hit of anti-pandemic will be no later than the end of April.\\
\noindent We can summarize our basic predictions as follows, for the worldwide and by country: 
\begin{itemize}
\item For the worldwide, (see Figure \ref{ww_conf} and Figure \ref{ww_dt}), overall, except in a period, the validation data show a well agreement with our forecast and all fall into the 97\% confidence interval (shaded area). 
We see the number of cases is almost equal to our predictions, showing each country of the whole world must take strict measures to stop the pandemic. At $\sim$April 07, 2020 we may obtain $>$ 1 million 200,000 confirmed cases (see Table \ref{futurconfcases}).
\item For China (see Figure \ref{ch}), based on optimistic estimation, the pandemic of COVID-19 would soon be ended within few weeks in China.
\item For Italy (see Figure \ref{it}) and Iran (see Figure \ref{ir}), the success of anti-pandemic will be no later than the middle of Avril. The situation in Italy and Iran are still very severe. We expect it will end up at the beginning of May.
\item For Senegal (see Figure \ref{sn}), we don't have enough data, because the first case appear in the beginning of March. And we are delighted to see cases are lower than our predictions, showing the nation wide anti-pandemic measures in Senegal come into play. 
\end{itemize}
\begin{rem}
In Figure \ref{ww_conf} to \ref{sn}, the dots (color black) are the real data, the blue line is $\hat{y}$, the {\bf shaded area represents the confidence interval (97\%)}.
\end{rem}
\noindent Finally, due to the inclusion of suspected cases with clinical diagnosis into confirmed cases (quarantined cases), we can see severe situation in some cities, which requires much closer attention.

\newpage
\section{Conclusion and perspectives}\label{ccl}
In this article, we propose the generalized SIR model and forecasting (machine learning) to analysis the coronavirus pandemic (COVID-19). 
The authors think that the part of identification could be deepen.  Indeed, one way should be to to consider in general  the 
identification  of systems. It  is the operation of determining the dynamic model of a process (system) from input / output measurements. Knowledge of the dynamic model is necessary for the design and implementation of an efficient regulatory system.\\ 
In other words, the identification of a real dynamical system (called object) is to characterize another system (called model), starting from the experimental knowledge of the inputs and outputs so as to obtain  an identity of behavior. In practice, the purpose of identification is generally to determine the conduct model, which can be used to simulate, control or regulate a process. This model can be physical (in the sense of analog or digital simulator and reduced model), or an abstract model (mathematical model, i.e. system of algebraic or differential equations (ODE or PDE)). Finally, note that some addressed questions in  parameter identification are  the recognizing of sets (domains, curves, etc.).\\
Finally, note that some addressed questions in parameter identification are  the recognition of sets (domains, curves, etc.). And a final remark is the identification of geodesics: 
In almost every epidemic, there is a phenomenon of super-propagation (super spreader)- when a person transmits an infection to a large number of people. And super spreader phenomena may  appear.
The role of super-spreaders it is the fact, noticed by Erdös \cite{ER}, that in a random graph there is always a small number of very connected people, through which almost all geodesics pass. How  could it be possible to identify these facts in this pandemic?\\
The stochastic aspect would be very interesting because we are not even sure to have for the case of Senegal all the data. In addition there are infected people who do not develop the disease and spread it (the case of some children as it seems) not to  mention the case of asymptomatics that are not monitored and are the cause of community transmission. There is also the probability of touching an infected object:  too much hazard in the transmission of the disease. Let us propose  among many possibilities the following stochastic SIR model:
\begin{equation}\label{stoeq}
\left\{ 
\begin{array}{ll}
\frac{dS}{dt}=-\beta IS + \sigma_1 dW_1\\[3mm]
\frac{dI}{dt}= \beta IS-\gamma I + \sigma_2 dW_2\\[3mm]
\frac{dR}{dt}=\gamma I
\end{array}\right.
\end{equation}
\begin{itemize}
\item $S$ is the number of individuals susceptible to be infected at time $t$.
\item $I$ is the number of both asymtotimatic and symptomatic infected individuals at time $t$.
\item $R$ is the number of recovered persons at time $t$.
\item The parameters $\beta$ and $\gamma$ are the transmission rate through exposure of the disease and the rate of recovering.
\item $\sigma_1$ and $\sigma_2$ are volatility rates that may  depend on the time $t$, the suspects and the infected. $W_1$ and  $W_2$ are classical Brownian motions.
\end{itemize}
 For this model, the  complete theoretical study could be an interesting part such as existence and uniqueness results and qualitative properties. Let us quote for the readers that there is a Cauchy Lipschitz Theorem  for (\ref{stoeq}), see for example \cite{LCE}.
 \begin{rem}
We would like to underline that the case of Senegal must be deepen in the sense that there are not yet enough data. And it should be better to look for getting more data and even consider a stochastic model to take into account the difficulties that can be met in the  information system to  collect  enough data.
\end{rem}

\subsection*{Acknowledgement}
The authors thanks the Non Linear Analysis, Geometry and Applications (NLAGA) Project for supporting this work (\url{http://nlaga-simons.ucad.sn}). 

%

\end{document}